\documentclass[english,aps,manuscript,showpacs,preprint,superscriptaddress]{revtex4-1}
\usepackage[T1]{fontenc}
\usepackage[latin9]{inputenc}
\usepackage{color}
\usepackage{amsmath}
\usepackage{amssymb}
\usepackage{esint}

\makeatletter
\@ifundefined{textcolor}{}
{%
 \definecolor{BLACK}{gray}{0}
 \definecolor{WHITE}{gray}{1}
 \definecolor{RED}{rgb}{1,0,0}
 \definecolor{GREEN}{rgb}{0,1,0}
 \definecolor{BLUE}{rgb}{0,0,1}
 \definecolor{CYAN}{cmyk}{1,0,0,0}
 \definecolor{MAGENTA}{cmyk}{0,1,0,0}
 \definecolor{YELLOW}{cmyk}{0,0,1,0}
}

\usepackage{babel}

\makeatother

\usepackage{babel}
\begin{document}

\title{{\normalsize{}Hamiltonian integration methods for Vlasov-Maxwell
equations}}

\author{Yang He}

\affiliation{School of Nuclear Science and Technology and Department of Modern
Physics, University of Science and Technology of China, Hefei, Anhui
230026, China}

\address{Key Laboratory of Geospace Environment, CAS, Hefei, Anhui 230026,
China}

\author{Hong Qin}

\affiliation{School of Nuclear Science and Technology and Department of Modern
Physics, University of Science and Technology of China, Hefei, Anhui
230026, China}

\affiliation{Plasma Physics Laboratory, Princeton University, Princeton, NJ 0854}

\author{Yajuan Sun}

\affiliation{LSEC, Academy of Mathematics and Systems Science, Chinese Academy
of Sciences, P.O.Box 2719, Beijing 100190, China}

\author{Jianyuan Xiao}

\affiliation{School of Nuclear Science and Technology and Department of Modern
Physics, University of Science and Technology of China, Hefei, Anhui
230026, China}

\address{Key Laboratory of Geospace Environment, CAS, Hefei, Anhui 230026,
China}

\author{Ruili Zhang}

\affiliation{School of Nuclear Science and Technology and Department of Modern
Physics, University of Science and Technology of China, Hefei, Anhui
230026, China}

\address{Key Laboratory of Geospace Environment, CAS, Hefei, Anhui 230026,
China}

\author{Jian Liu}

\affiliation{School of Nuclear Science and Technology and Department of Modern
Physics, University of Science and Technology of China, Hefei, Anhui
230026, China}

\address{Key Laboratory of Geospace Environment, CAS, Hefei, Anhui 230026,
China}
\begin{abstract}
{\normalsize{}Hamiltonian integration methods for the Vlasov-Maxwell
equations are developed by a Hamiltonian splitting technique. The
Hamiltonian functional is split into five parts, i.e., the electrical
energy, the magnetic energy, and the kinetic energy in three Cartesian
components. Each of the subsystems is a Hamiltonian system with respect
to the Morrison-Marsden-Weinstein Poisson bracket and can be solved
exactly. Compositions of the exact solutions yield Poisson structure
preserving, or Hamiltonian, integration methods for the Vlasov-Maxwell
equations, which have superior long-term fidelity and accuracy.}{\normalsize \par}
\end{abstract}

\keywords{Structure preserving algorithm, Poisson structure, Vlasov-Maxwell
equations}

\maketitle
The dynamics of charged particles in a plasma interacting with the
self-consistent electromagnetic fields can be described by the Vlasov-Maxwell
(VM) equations. In modern plasma physics and accelerator physics,
numerical integration of the Vlasov-Maxwell equations is an important
tool for theoretical studies, and varieties of numerical algorithms
have been developed. Recently, geometric integration methods \citep{Ruth83,Feng85,Feng10,Hairer03},
which are designed in the spirit of preserving the intrinsic structures
of a dynamical system, have been developed for plasma physics applications
\citep{Qin08,Qin09,Qin13,Zhang15,He15,Squire12,Xiao13,Kraus14,Burby14-1,Zhou14,Ellison15,Shadwick14,Evs13,crous15hamiltonian,Qin15}.
By preserving properties such as the Poisson structure of a Hamiltonian
system and the invariant volume form of a source-free system, geometric
integration methods usually generate numerical results with superior
long-term behavior compared to other methods \citep{Hairer03,shangk99kto},
and are thus more suitable for\textcolor{red}{{} }large-scale, long-term
simulations. It is known that the Vlasov-Maxwell system is a Hamiltonian
system with respect to a Poisson bracket \citep{Morrison80Poisson,Morrison81Poisson,Marsden82VM}.
In a recent paper \citep{crous15hamiltonian}, Crouseilles, Einkemmer,
and Faou proposed an innovative Hamiltonian splitting method for the
Vlasov-Maxwell equations based on a bracket first suggested in Ref.\,\citep{Morrison80Poisson}.
This splitting scheme results in three solvable subsystems, whose
exact solutions can be combined into algorithms for the Vlasov-Maxwell
equations that preserve the structure specified by the bracket. However,
it has been pointed out in Ref.\,\citep{qin15comment} that the bracket
adopted in Ref.\,\citep{crous15hamiltonian} is not Poisson, because
it does not satisfy the Jacobi identity \citep{Morrison81Poisson,Marsden82VM}.
Very disappointedly, if the Hamiltonian splitting method proposed
in Ref. \citep{crous15hamiltonian} is applied with the correct Poisson
bracket, a.k.a. the Morrison-Marsden-Weinstein (MMW) bracket \citep{Morrison81Poisson,Marsden82VM,Burby14,qin15comment},
one of the subsystems cannot be solved exactly in general. Therefore,
Hamiltonian integration methods for the Vlasov-Maxwell equations cannot
be constructed by using the splitting method developed in Ref.\,\citep{crous15hamiltonian}.

In this paper, we apply the MMW Poisson bracket \citep{Morrison81Poisson,Marsden82VM}
and develop a family of Hamiltonian integration methods via a new
splitting of the Hamiltonian functional. In the current context, a
Hamiltonian integration method is defined as a numerical integrator
that preserves the Poisson structure of the Vlasov-Maxwell system
specified by the MMW bracket. In addition to splitting the Hamiltonian
into electrical energy, magnetic energy, and kinetic energy, we further
split the kinetic energy into three Cartesian components. It turns
out that exact solutions for all of the resulted subsystems can be
calculated by the method of characteristics. By compositions of the
exact solutions to the subsystems, methods of theoretically arbitrarily
high order in time for the original Vlasov-Maxwell equations can be
constructed. The exact solutions of the subsystems preserve the same
Poisson structure as the Vlasov-Maxwell equations, so do the combined
algorithms.\textcolor{red}{{} }Therefore, the good properties of a Hamiltonian
integrator, such as the long-term stability and accuracy and global
bound on the energy error, are all inherited by this family of new
integration methods for the Vlasov-Maxwell system.

The Vlasov-Maxwell equations considered in the present study are
\begin{gather}
\frac{\partial f}{\partial t}+{\bf v}\cdot\frac{\partial f}{\partial{\bf x}}+\left({\bf E}+{\bf v}\times{\bf B}\right)\cdot\frac{\partial f}{\partial{\bf v}}=0,\label{eq:Vlasov}\\
\nabla\times{\bf B}=\int{\bf v}f({\bf x},{\bf v},t)d{\bf v}+\frac{\partial{\bf E}}{\partial t},\label{eq:MaxwellE}\\
\nabla\times{\bf E}=-\frac{\partial{\bf B}}{\partial t},\label{eq:MaxwellB}\\
\nabla\cdot{\bf E}=\int f({\bf x},{\bf v},t)d{\bf v},\label{eq:divE}\\
\nabla\cdot{\bf B}=0,\label{eq:divB}
\end{gather}
where $f({\bf x},{\bf v},t)$ is the distribution function of position
${\bf x}\in U\subset\mathbb{R}^{3}$ and velocity ${\bf v}\in\mathbb{R}^{3}$
at time $t$, and $({\bf E}({\bf x},t),{\bf B}({\bf x},t))\in\mathbb{R}^{3}\times\mathbb{R}^{3}$
are the electromagnetic fields. For easy presentation, the species
index, charge, mass, and other constant are omitted. Equations (\ref{eq:Vlasov})-(\ref{eq:MaxwellB})
are closed, and equations (\ref{eq:divE})-(\ref{eq:divB}) result
from the gauge symmetry of the system. According to \cite{Marsden82VM},
the VM equations (\ref{eq:Vlasov})-(\ref{eq:divB}) are equivalent
to the Hamiltonian system
\begin{equation}
\frac{\partial\mathcal{F}}{\partial t}=\{\{\mathcal{F},\mathcal{H}\}\}\label{eq:PoissonVM}
\end{equation}
on the phase space $\mathcal{MV}=\left\{ ({\bf E},{\bf B})|\nabla\cdot{\bf B}=0,\nabla\cdot{\bf E}=\int fd{\bf v}\right\} $,
with $\mathcal{F}(f,{\bf E},{\bf B})$ being any functional on $\mathcal{MV}$
and $\mathcal{H}$ being the Hamiltonian functional defined as
\begin{equation}
\mathcal{H}(f,\mathbf{E},\mathbf{B})=\frac{1}{2}\int|\mathbf{v}|{}^{2}f({\bf x},{\bf v},t)d\mathbf{x}d\mathbf{v}+\frac{1}{2}\int\left(|\mathbf{E}({\bf x},t)|^{2}+|\mathbf{B}({\bf x},t)|^{2}\right)d\mathbf{x}.\label{eq:HamiltonVM}
\end{equation}
Here, $\{\{\cdot,\cdot\}\}$ denotes the Morrison-Marsden-Weinstein
bracket \citep{Marsden82VM,Morrison80Poisson,Morrison81Poisson},
\begin{align}
\{\{\mathcal{F},\mathcal{G}\}\} & (f,\mathbf{E},\mathbf{B})=\int f\left\{ \frac{\delta\mathcal{F}}{\delta f},\frac{\delta\mathcal{G}}{\delta f}\right\} _{\mathbf{xv}}d\mathbf{x}d\mathbf{v}\nonumber \\
 & +\int\left[\frac{\delta\mathcal{F}}{\delta\mathbf{E}}\cdot\left(\triangledown\times\frac{\delta\mathcal{G}}{\delta\mathbf{B}}\right)-\frac{\delta\mathcal{G}}{\delta\mathbf{E}}\cdot\left(\triangledown\times\frac{\delta\mathcal{F}}{\delta\mathbf{B}}\right)\right]d\mathbf{x}\label{eq:MMWB}\\
 & +\int\left(\frac{\delta\mathcal{F}}{\delta\mathbf{E}}\cdot\frac{\partial f}{\partial\mathbf{v}}\frac{\delta\mathcal{G}}{\delta f}-\frac{\delta\mathcal{G}}{\delta\mathbf{E}}\cdot\frac{\partial f}{\partial\mathbf{v}}\frac{\delta\mathcal{F}}{\delta f}\right)d\mathbf{x}d\mathbf{v}\nonumber \\
 & +\int f{\bf B}\cdot\left(\frac{\partial}{\partial\mathbf{v}}\frac{\delta\mathcal{F}}{\delta f}\times\frac{\partial}{\partial\mathbf{v}}\frac{\delta\mathcal{G}}{\delta f}\right)d\mathbf{x}d\mathbf{v}.\nonumber
\end{align}
In the first term on the right-hand side, $\left\{ \cdot,\cdot\right\} _{\mathbf{xv}}$
denotes the canonical Poisson bracket for functions of $\left({\bf x},{\bf v}\right)$.
With the initial conditions $f({\bf x},{\bf v},0)=f_{0}({\bf x},{\bf v})$
and $({\bf E}({\bf x},0),{\bf B}({\bf x},0))\in\mathcal{MV}$, there
exists unique solution to the system (\ref{eq:PoissonVM}), on which
the bracket (\ref{eq:MMWB}) is Poisson and is preserved. Given this
Hamiltonian formulation of the Vlasov-Maxwell system, Poisson structure
preserving integration methods can be constructed as follows. The
system is first split into several solvable subsystems by decomposing
the Hamiltonian functional (\ref{eq:HamiltonVM}). We then find the
exact solutions to the subsystems, and finally the exact solutions
of the subsystems are composed in a proper way to construct integrators
for the Vlasov-Maxwell equations that preserve the Poisson structure
(\ref{eq:MMWB}).

Firstly, we follow Ref. \citep{crous15hamiltonian} to split the Hamiltonian
into three parts,
\begin{equation}
\begin{gathered}\mathcal{H}=\mathcal{H}_{E}+\mathcal{H}_{B}+\mathcal{H}_{f},\\
\mathcal{H}_{E}=\frac{1}{2}\int|\mathbf{E}({\bf x},t)|^{2}d\mathbf{x},\quad\mathcal{H}_{B}=\frac{1}{2}\int|\mathbf{B}({\bf x},t)|^{2}d\mathbf{x},\quad\mathcal{H}_{f}=\frac{1}{2}\int|\mathbf{v}|{}^{2}f({\bf x},{\bf v},t)d\mathbf{x}d\mathbf{v}.
\end{gathered}
\label{eq:HEB}
\end{equation}
Using the MMW bracket (\ref{eq:MMWB}) and the Hamiltonian equation
(\ref{eq:PoissonVM}), the VM equations (\ref{eq:Vlasov})-(\ref{eq:divB})
can be split into three subsystems on $\mathcal{MV}$,
\[
\dot{\mathcal{F}}=\left\{ \left\{ \mathcal{F},\mathcal{H}_{E}\right\} \right\} ,\quad\dot{\mathcal{F}}=\left\{ \left\{ \mathcal{F},\mathcal{H}_{B}\right\} \right\} ,\quad\dot{\mathcal{F}}=\left\{ \left\{ \mathcal{F},\mathcal{H}_{f}\right\} \right\} .
\]

Next, we try to solve the subsystems for exact solutions. The subsystem
$\dot{\mathcal{F}}=\left\{ \left\{ \mathcal{F},\mathcal{H}_{E}\right\} \right\} $
associated with the Hamiltonian $\mathcal{H}_{E}$ is equivalent to
\begin{equation}
\begin{gathered}\frac{\partial f}{\partial t}+\mathbf{E}(\mathbf{x},t)\cdot\frac{\partial f}{\partial{\bf v}}=0,\\
\frac{\partial\mathbf{E}}{\partial t}=0,\\
\frac{\partial\mathbf{B}}{\partial t}=-\nabla\times\mathbf{E}.
\end{gathered}
\label{eq:VMHE}
\end{equation}
Given the initial functions $f_{0}$ and $({\bf E}({\bf x},{0}),{\bf B}({\bf x},{0}))\in\mathcal{MV}$
, the solution to the subsystem (\ref{eq:VMHE}) is
\begin{equation}
\begin{gathered}f(\mathbf{x},\mathbf{v},t)=f_{0}(\mathbf{x},\mathbf{v}-t\mathbf{E}(\mathbf{x},0)),\\
\mathbf{E}(\mathbf{x},t)=\mathbf{E}(\mathbf{x},0),\\
\mathbf{B}(\mathbf{x},t)=\mathbf{B}(\mathbf{x},0)-t\nabla\times\mathbf{E}(\mathbf{x},0).\text{}
\end{gathered}
\label{eq:SoHE}
\end{equation}
Using the notation $\exp(\mathcal{H}_{E}t)$ as the update operator
in time,\textcolor{red}{{} }we denote this solution formally as
\[
(f({\bf x,v},t),\mathbf{E}({\bf x},t),\mathbf{B}({\bf x},t))^{T}=\exp(\mathcal{H}_{E}t)(f_{0}({\bf x,v}),\mathbf{E}(\mathbf{x},0),\mathbf{B}(\mathbf{x},0))^{T}.
\]
The subsystem $\dot{\mathcal{F}}=\left\{ \left\{ \mathcal{F},\mathcal{H}_{B}\right\} \right\} $
corresponding to the Hamiltonian $\mathcal{H}_{B}$ is equivalent
to,
\begin{equation}
\begin{gathered}\frac{\partial f}{\partial t}=0,\\
\frac{\partial\mathbf{E}}{\partial t}=\nabla\times\mathbf{B},\\
\frac{\partial\mathbf{B}}{\partial t}=0.
\end{gathered}
\label{eq:VMHB}
\end{equation}
With initial conditions given on $\mathcal{MV}$, the solution to
the subsystem (\ref{eq:VMHB}) is
\begin{equation}
\begin{gathered}f(\mathbf{x},\mathbf{v},t)=f_{0}(\mathbf{x},\mathbf{v}),\\
\mathbf{E}(\mathbf{x},t)=\mathbf{E}(\mathbf{x},0)+t\nabla\times\mathbf{B}(\mathbf{x},0),\\
\mathbf{B}(\mathbf{x},t)=\mathbf{B}(\mathbf{x},0).\text{}
\end{gathered}
\label{eq:SoHB}
\end{equation}
We denote the corresponding update operator as $\exp(\mathcal{H}_{B}t)$.

For the Hamiltonian $\mathcal{H}_{f}$, however, the subsystem $\dot{\mathcal{F}}=\left\{ \left\{ \mathcal{F},\mathcal{H}_{f}\right\} \right\} $
is in a more complicated form,
\begin{equation}
\begin{gathered}\frac{\partial f}{\partial t}+\mathbf{v}\cdot\frac{\partial f}{\partial{\bf x}}+(\mathbf{v}\times\mathbf{B}({\bf x},t))\cdot\frac{\partial f}{\partial{\bf v}}=0,\\
\frac{\partial\mathbf{E}}{\partial t}=-\int\mathbf{v}f(\mathbf{x},\mathbf{v},t)d\mathbf{v},\\
\frac{\partial\mathbf{B}}{\partial t}=0.
\end{gathered}
\label{eq:VMHf}
\end{equation}
It has been shown that unless the magnetic field ${\bf B}$ vanishes
or is uniform in space, this system can not be solved exactly \citep{qin15comment}.
Therefore, we search for a Poisson structure preserving method for
Eq.\,(\ref{eq:VMHf}) by further splitting the Hamiltonian $\mathcal{H}_{f}$
into more solvable subsystems. Utilizing the structure of the cross
product in the Cartesian frame,
\[
\mathbf{v}\times\mathbf{B}=\mathbf{\hat{B}v}=\left[\begin{array}{ccc}
0 & B_{3} & -B_{2}\\
-B_{3} & 0 & B_{1}\\
B_{2} & -B_{1} & 0
\end{array}\right]\mathbf{v},
\]
we further split the Hamiltonian $\mathcal{H}_{f}$ into different
Cartesian components,\textcolor{red}{{} }i.e.,
\begin{equation}
\mathcal{H}_{f}=\mathcal{H}_{1f}+\mathcal{H}_{2f}+\mathcal{H}_{3f},\quad\mathcal{H}_{if}=\frac{1}{2}\int v_{i}^{2}f(\mathbf{x},\mathbf{v},t)d\mathbf{x}d\mathbf{v},\quad i=1,2,3.\label{eq:Hf}
\end{equation}
The subscript of $B$ and $v$ denotes the corresponding Cartesian
component of the vector. The subsystem associated with each Hamiltonian
$\mathcal{H}_{if}$ is
\begin{equation}
\begin{gathered}\frac{\partial f}{\partial t}+v_{i}\frac{\partial f}{\partial x_{i}}-B_{i-1}(\mathbf{x})v_{i}\frac{\partial f}{\partial v_{i+1}}+B_{i+1}(\mathbf{x})v_{i}\frac{\partial f}{\partial v_{i-1}}=0,\\
\frac{\partial E_{i}}{\partial t}=-\int v_{i}f(\mathbf{x},\mathbf{v},t)d\mathbf{v},\\
\frac{\partial\mathbf{B}}{\partial t}=0,\quad\frac{\partial E_{i-1}}{\partial t}=\frac{\partial E_{i+1}}{\partial t}=0,
\end{gathered}
\label{eq:VMHfi}
\end{equation}
with $v_{4}:=v_{1}$ and $v_{0}:=v_{3}$. The first equation of (\ref{eq:VMHfi})
can be solved by the method of characteristics, and the characteristic
equations are
\[
\begin{aligned}\dot{x}_{i}=v_{i},\quad\dot{v}_{i+1}=-B_{i-1}(\mathbf{x})v_{i},\quad\dot{v}_{i-1}=B_{i+1}(\mathbf{x})v_{i.}\end{aligned}
\]
It is a system of ordinary differential equations, whose exact solution
is
\[
\begin{gathered}x_{i}(t)=x_{i}(0)+tv_{i},\\
v_{i+1}(t)%
=v_{i+1}(0)-\int_{x_{i}(0)}^{x_{i}(t)}B_{i-1}({\bf x})dx_{i},\\
v_{i-1}(t)%
=v_{i-1}(0)+\int_{x_{i}(0)}^{x_{i}(t)}B_{i+1}({\bf x})dx_{i}.
\end{gathered}
\]
Therefore, for the initial conditions $f({\bf x},{\bf v},0)=f_{0}({\bf x,v})$
and $\left({\bf E}(\mathbf{x},0),{\bf B}(\mathbf{x},0)\right)\in\mathcal{MV}$,
the exact solution to the subsystem (\ref{eq:VMHfi}) is
\begin{equation}
\begin{gathered}f(\mathbf{x},\mathbf{v},t)=f_{0}({\bf x}-tv_{i}{\bf e}_{i},{\bf v}+F_{i-1}{\bf e}_{i+1}-F_{i+1}{\bf e}_{i-1}),\\
F_{l}=\int_{x_{i}-tv_{i}}^{x_{i}}B_{l}({\bf x})dx_{i},\quad l=i+1,i-1,\\
{\bf E}({\bf x},t)={\bf E}({\bf x},0)-\int_{0}^{t}\int{\bf e}_{i}v_{i}f({\bf x},{\bf v},\tau)d{\bf v}d\tau,\\
{\bf B}(\mathbf{x},t)={\bf B}(\mathbf{x},0),\text{},
\end{gathered}
\label{eq:soHfi}
\end{equation}
where ${\bf e}_{i}$ is the unit vector in the $i$-th Cartesian direction.
We denote the exact solution given by Eq.\,(\ref{eq:soHfi}) as $(f,{\bf E},{\bf B})(t)^{\mathtt{T}}=\exp(\mathcal{H}_{if}t)(f_{0}({\bf x},{\bf v}),{\bf E}({\bf x},0),{\bf B}({\bf x},0))^{\mathtt{T}}$.
More specifically, the exact solution of the subsystem corresponding
to $\mathcal{H}_{1f}$ is
\[
\begin{gathered}f({\bf x},{\bf v},t)=f_{0}(x_{1}-tv_{1},x_{2},x_{3},v_{1},v_{2}+v_{1}F_{3},v_{3}-v_{1}F_{2}),\\
F_{l}=\int_{x_{1}-tv_{1}}^{x_{1}}B_{l}({\bf x})dx_{1},l=2,3,\\
E_{1}({\bf x},t)=E_{1}({\bf x},0)-\int_{0}^{t}\int v_{1}f({\bf x},{\bf v},\tau)d{\bf v}d\tau.
\end{gathered}
\]
The exact solution associated with $\exp(\mathcal{H}_{2f}t)$ is
\[
\begin{gathered}f({\bf x},{\bf v},t)=f_{0}(x_{1},x_{2}-tv_{2},x_{3},v_{1}-v_{2}F_{3},v_{2},v_{3}+v_{2}F_{1}),\\
F_{l}=\int_{x_{2}-tv_{2}}^{x_{2}}B_{l}({\bf x})dx_{2},l=1,3,\\
E_{2}({\bf x},t)=E_{2}({\bf x},0)-\int_{0}^{t}\int v_{2}f({\bf x},{\bf v},\tau)d{\bf v}d\tau,
\end{gathered}
\]
and the exact solution associated with $\exp(\mathcal{H}_{3f}t)$
is
\[
\begin{gathered}f({\bf x},{\bf v},t)=f_{0}(x_{1},x_{2},x_{3}-tv_{3},v_{1}+v_{3}F_{2},v_{2}-v_{3}F_{1},v_{3}),\\
F_{l}=\int_{x_{3}-tv_{3}}^{x_{3}}B_{l}({\bf x})dx_{3},l=1,2,\\
E_{3}({\bf x},t)=E_{3}({\bf x},0)-\int_{0}^{t}\int v_{3}f({\bf x},{\bf v},\tau)d\mathbf{v}d\tau.
\end{gathered}
\]
Overall, we split the Vlasov-Maxwell equations into five subsystems,
\begin{gather*}
\mathcal{H}=\mathcal{H}_{E}+\mathcal{H}_{B}+\mathcal{H}_{1f}+\mathcal{H}_{2f}+\mathcal{H}_{3f},\\
\dot{\mathcal{F}}=\left\{ \left\{ \mathcal{F},\mathcal{H}_{E}\right\} \right\} ,\thinspace\thinspace\dot{\mathcal{F}}=\left\{ \left\{ \mathcal{F},\mathcal{H}_{B}\right\} \right\} ,\thinspace\thinspace\dot{\mathcal{F}}=\left\{ \left\{ \mathcal{F},\mathcal{H}_{1f}\right\} \right\} ,\thinspace\thinspace\dot{\mathcal{F}}=\left\{ \left\{ \mathcal{F},\mathcal{H}_{2f}\right\} \right\} ,\thinspace\thinspace\dot{\mathcal{F}}=\left\{ \left\{ \mathcal{F},\mathcal{H}_{3f}\right\} \right\} .
\end{gather*}
with $\mathcal{H}_{E},\mathcal{H}_{B}$ and $\mathcal{H}_{if},\thinspace i=1,2,3,$
defined in Eqs.\,(\ref{eq:HEB}) and (\ref{eq:Hf}), and the Poisson
bracket being defined in Eq.\,(\ref{eq:MMWB}). The exact solutions
to the subsystems are given explicitly by Eqs.\,(\ref{eq:SoHE}),
(\ref{eq:SoHB}), and (\ref{eq:soHfi}), respectively. For initial
functions defined on $\mathcal{MV}$, the solutions of the subsystems
are all on $\mathcal{MV}$ and preserve the MMW Poisson bracket.

Given exact solutions to the subsystems, integration methods for the
original Vlasov-Maxwell equations can be constructed by compositions.
For example, if we denote the solution at time $t$ as $Z(t):=(f({\bf x,{\bf v}},t),\mathbf{E}({\bf x},t),\mathbf{B}({\bf x},t))^{T}$,
a first order numerical update in one time step with step size $\Delta t$
can be derived from the Lie-Trotter composition
\[
Z(t+\Delta t)=\exp(\Delta t\mathcal{H}_{E})\exp(\Delta t\mathcal{H}_{B})\exp(\Delta t\mathcal{H}_{1f})\exp(\Delta t\mathcal{H}_{2f})\exp(\Delta t\mathcal{H}_{3f})Z(t),
\]
and a second order symmetric method can be constructed by the following
symmetric composition,
\begin{eqnarray*}
Z(t+\Delta t)= &  & \exp\left(\frac{\Delta t}{2}\mathcal{H}_{E}\right)\exp\left(\frac{\Delta t}{2}\mathcal{H}_{B}\right)\exp\left(\frac{\Delta t}{2}\mathcal{H}_{1f}\right)\exp\left(\frac{\Delta t}{2}\mathcal{H}_{2f}\right)\exp\left(\Delta t\mathcal{H}_{3f}\right)\text{}\\
 &  & \exp\left(\frac{\Delta t}{2}\mathcal{H}_{2f}\right)\exp\left(\frac{\Delta t}{2}\mathcal{H}_{1f}\right)\exp\left(\frac{\Delta t}{2}\mathcal{H}_{B}\right)\exp\left(\frac{\Delta t}{2}\mathcal{H}_{E}\right)Z(t).
\end{eqnarray*}
With the help of the BCH formula, proper compositions can be found
to yield methods of arbitrarily high order in time \citep{mclachlanq02,Hairer03}.
Given the geometrical properties of the solutions to the subsystems,
the combined methods will preserve the Poisson structure and generate
solutions on $\mathcal{MV}$,\textcolor{red}{{} }if the initial values
$Z(0)$ are taken on $\mathcal{MV}$.

In conclusion, we have constructed a family of Poisson structure preserving,
or Hamiltonian, integration methods for the Vlasov-Maxwell equations
by the splitting technique. The Hamiltonian is split into five parts,
each part associates with a solvable Hamiltonian subsystem with respect
to the MMW Poisson bracket. As a consequence, integration methods
for the Vlasov-Maxwell equations constructed via composition of the
exact solutions of the subsystems will preserve the original Poisson
structure. These new Hamiltonian methods are expected to exhibit long-term
accuracy and fidelity, as well as bounded error in energy and other
invariants. Numerical applications of the methods will be reported
in future publications.

In the present study, we have focused on the time integration of the
Vlasov-Maxwell system and the preservation of the original Poisson
structure. If the system is discretized in space, the resulted scheme
needs to preserve a discrete Poisson bracket depending on the spatial
discretization scheme \cite{Qin15}.

As a final note, we emphasize that this general methodology of constructing
Hamiltonian integration algorithms via Hamiltonian splitting is also
applicable to other systems that admit Poisson structures. Recently,
Burby et al. discovered the Poisson structures for the gyrokinetic
system \cite{Burby14} and the collision operator \cite{Burby15}.
Application of the splitting technique to these systems will generate
more effective numerical algorithms for large-scale, long-term simulation
studies of plasma physics.
\begin{acknowledgments}
This research was supported by the National Natural Science Foundation
of China (11271357, 11261140328, 11305171), the CAS Program for Interdisciplinary
Collaboration Team, the ITER-China Program (2015GB111003, 2014GB124005,
2013GB111000), the JSPS-NRF-NSFC A3 Foresight Program in the field
of Plasma Physics (NSFC-11261140328), and the Fundamental Research
Funds for the Central Universities (WK2030040057).\end{acknowledgments}

\end{document}